\documentclass[aps,prl,twocolumn,showpacs]{revtex4}

\usepackage{amsmath}
\usepackage{epstopdf}
\usepackage{color}

\usepackage{epsfig}
\usepackage{graphicx}
 
\usepackage{amssymb}
\usepackage{amsfonts}
\usepackage{dcolumn}

\begin{document}


\title{Symmetry breaking and multi-hump solitons in inhomogeneous gain landscapes}

\author{Yaroslav V.  Kartashov$^1$, Vladimir V. Konotop$^2$, and Victor A. Vysloukh$^1$}

\affiliation{$^1$ICFO-Institut de Ciencies Fotoniques, and Universitat Politecnica de Catalunya, Mediterranean Technology Park, 08860 Castelldefels (Barcelona), Spain
 \\
$^2$Centro de F\'{\i}sica Te\'orica e Computacional
  and Departamento de F\'{\i}sica, Faculdade de Ci\^encias, Universidade de Lisboa,
Avenida Professor Gama
Pinto 2, Lisboa 1649-003, Portugal
}

\begin{abstract}
\par

We address one-dimensional soliton formation in the cubic nonlinear medium with two-photon absorption and transversally inhomogeneous gain landscape consisting of a single or several amplifying channels.  Existence of the solitons  requires certain threshold gain 
while the properties of solitons strongly depend on whether the number of the amplifying channels is odd or even. In the former case increase of the gain leads to a symmetry breaking, which occurs through the pitchfork bifurcation, and to emergence of a single or several co-existing stable asymmetric modes. In the case of even number of amplifying channels we have found only asymmetric stable states.

\end{abstract}



\maketitle

Emergence of localized nonlinear patterns supported by  localized gain recently attracted increasing attention. There have been reported stable one-dimensional  (1D) structures in media described by the cubic complex Guinzburg-Landau equation with   linear losses and either one~\cite{Malomed} or two~\cite{Malomed_two} highly localized  "hot spots". Stable solitons were also found in periodic lattices with   active channels and nonlinear two-photon absorption~\cite{KKVT}, and at the surface of a periodic medium~\cite{KKV}. Stable patterns may exist even in 2D 
settings, in layered structures of the planar waveguides~\cite{Kutz} and in 2D
lattices~\cite{KKVT_2D}.

Localized gain significantly changes the physics of emergent nonlinear patterns. In addition to standard constraints imposed by the balance between the dissipation and gain, it introduces a new spatial scale -- the width of the gain domain. This suggests a possibility of existence of more sophisticated 
structures, than the simplest symmetric and/or anti-symmetric 
dissipative  solitons. 
In particular, 
it is of fundamental interest to exploit the phenomenon of the symmetry breaking, which is known to be generic for conservative nonlinear systems possessing a characteristic spatial scale.  These are for example,  systems governed by the  nonlinear Schr\"odinger   equation  with a symmetric double-well potential~\cite{double-well,Panos}. Self-trapping in one of the two identical channels,  
has already been  observed experimentally in  nonlinear optics~\cite{sym_break_opt} and in a Bose-Einstein condensate~\cite{BEC}. Symmetry breaking in a dual-core dissipative fiber with  parametric gain and linear losses, which occurs due to interplay of the nonlinearity and coupling between two cores defined by the modulation of the linear refractive index,
 was reported in~\cite{com}.
Asymmetric dissipative solitons were reported 
only in 2D  systems, e.g. in cubic-quintic Ginzburg-Landau equation~\cite{Nail}  and in media with saturable gain and absorption~\cite{Rosanov}.   
Here we report a principally different scenario of the symmetry breaking. It  occurs in a medium without any 
conservative potential or modulation of the refractive index, but having nonlinear dissipation and localized gain.

Our setting is related to   dissipative solitons, observed  at  the wavelength 1319 nm in  self-focusing electrically pumped waveguides 
fabricated on an InP substrate~\cite{exper}. In such structures the two-photon absorption   usually ranges from $10^{-1}$ to $10^{-2}$ cm/GW while localized gain can be implemented   by using segmented strip-like electrodes or spatially-localized optical pumping. Solitons can be excited in 0.6$\,\mu$m-thick planar guiding layers with a length of several millimeters, by input beams with typical waists of a few micrometers, at gain levels about 70 cm$^{-1}$~\cite{exper}.    
  The considered model is also relevant for description of Bose-Einstein condensates of quasiparticles
in presence of nonresonant pumping~\cite{polariton}. 

We consider the propagation of laser radiation in a focusing cubic medium with strong two-photon absorption and transversally inhomogeneous gain 
described by the equation for the dimensionless light field amplitude $q$:
\begin{eqnarray}
i\frac{\partial q}{\partial \xi} = - \frac 12 \frac{\partial^2 q}{\partial \eta^2} +i p_iR(\eta)q -  |q|^2q - i\alpha |q|^2q
\label{NLS}
\end{eqnarray}
where  $\xi$ and $\eta$ are the normalized longitudinal and  transverse    coordinates, respectively; $p_i>0$  is the gain parameter; $R(\eta)$   describes the transverse gain profile (its amplitude is normalized to one); $\alpha>0$  is the strength of two-photon absorption.  
We consider gain landscapes containing an integer number $n$ of periods of $\cos^2\eta$. For example, to model a gain with an odd number of amplifying channels we set  $R(\eta)=\cos^2\eta$ for $|\eta|\leq \eta_n $, where $\eta_n=n\pi/2$, and $R(\eta)=0$  for $|\eta|>\eta_n$, and  vary  $p_i$, $\alpha$ and $n$.
Assuming the characteristic transverse scale to be $3\,\mu$m, we estimate the longitudinal scale (the diffraction length) to be  $\sim 170\,\mu$m at the wavelength of 1.32$\,\mu$m. Then $p_i=1$ corresponds to the linear gain  $\sim 60\, \mbox{cm}^{-1}$  and $\alpha=1$ corresponds to the two-photon absorption coefficient $\approx\,$0.017 cm/GW. The linear absorption is supposed to be compensated by the   gain  for $|\eta|>\eta_n$.

Dissipative solitons  of Eq.~(\ref{NLS}) can be searched in the form $\psi=w(\eta)e^{ib\xi}$, where $b$ is the propagation constant, $w(\eta)=w_r+iw_i=u(\eta)e^{i\theta(\eta)}$ is a complex amplitude, with real, $w_r$, and imaginary, $w_i$, parts. The modulus $u$ and phase $\theta$ solve the equations  
\begin{eqnarray}
\label{rho}
bu=\frac{u_{\eta\eta}}{2} 
+ u^3 -\frac{j^2}{2u^3},\quad
j_\eta 
=2p_iR(\eta)u^2-2\alpha u^4.
\end{eqnarray}
where $j(\eta)\equiv\theta_\eta u^2$, can be referred to as a current density. We are interested in  localized solutions with   $ u, j \to 0$  at $|\eta|\to\infty$, which can be obtained for $b>0$ (more specifically with the exponentially decaying asymptotics $\displaystyle{u\sim e^{-\sqrt{2b}|\eta|}}$ and $\displaystyle{j\sim e^{-4\sqrt{2b}|\eta|}}$).

Localized modes of the system~(\ref{rho}) form if a focusing nonlinearity counterbalances diffractive broadening, i.e. when the following equation 
\begin{eqnarray}
\label{diff}
bU=-\frac 12\int u_\eta ^2d\eta-\int\frac{j^2}{u^2}d\eta+\int u^4d\eta
\end{eqnarray}
where $U=\int u^2d\eta$ is the energy flow, is satisfied and when the nonlinear losses integrally compensate the spatially inhomogeneous gain, i.e. when
\begin{eqnarray}
\label{pump}
p_i\int R(\eta)u^2d\eta=\alpha \int u^4d\eta
\end{eqnarray}

The above formulas allow one to argue on possibility of existing of two different types of the   modes at the same parameters of the system. Consider  the limit of high amplification  $p_i\to \infty$. Assuming that  the solution amplitude ${\cal A}=\max\{u\}$ grows and the width $\ell$  decreases, the  relation 
(\ref{pump})  
suggests the scaling ${\cal A}\sim 1/\ell\sim \sqrt{p_i/\alpha}$. 
This allows us to deduce from (\ref{pump}) the estimate $U\approx (\alpha/p_i)\int u^4d\eta\sim\sqrt{p_i/\alpha}$,  valid subject to the assumption that the soliton is maximum   is placed exactly  
 at $\eta=0$ where the pump has the maximal value, i.e.   valid for a symmetric mode.  
For a crude  guess of the proportionality coefficient  
in this estimate we use the ansatz  $u\approx {\nu}/{\cosh(\nu\eta)}$ [that corresponds to neglecting the current $j$ in the first of equations (\ref{rho}),  what  strictly speaking  can be done only in the vicinity of $\eta=0$].   This, allows us to obtain from (\ref{pump})  for the symmetric mode 
$\nu=\sqrt{{3p_i}/{2\alpha}}$ and thus    $U\sim 2\nu=\sqrt{6p_i/\alpha}$ and $b\sim \nu^2/2= 3p_i/4\alpha$. 

However constraint (\ref{pump}) admits yet another scaling  where a solution   width  grows with $p_i$.  Then for wide solutions, i.e. at $\ell\gg \eta_0=\pi$, from  Eq.~(\ref{pump}) the relation $p_i\sim\alpha{\cal A}^2\ell$ follows. On the other hand, now $U\sim {\cal A}^2\ell$, i.e. $U\sim p_i/\alpha$. Thus, unlike in the previous estimate, now we are  restricted neither by the position of the maximum of the mode, nor by   the symmetry of its shape. Moreover, in the corresponding solution, the diffraction term $u_{\eta\eta}\sim{\cal A}/\ell^2$ cannot be compensated  by the Kerr nonlinearity $u^3\sim{\cal A}^3$ alone, and the role of the current distribution, i.e. of $j^2/u^3$,  becomes crucial (it reduces the impact of the Kerr nonlinearity).
Notice   that the major influence of the current occurs not at the origin 
(where for the symmetric solutions it is exactly zero) 
but at some intermediate point $\eta_*$ defined by the the condition $j_\eta(\eta_*)=0$. Thus if a solution  with the suggested scaling exists, it should have asymmetric shape, with the maximum located in the vicinity of the point $\eta_*$ (at least in the limit $p_i\to \infty$).  

Further information about  the maximal field amplitude ${\cal A}$ can be obtained from Eq.~(\ref{rho}). Indeed, for $\eta>\eta_n$ the current is decaying, 
  $j_\eta=-2\alpha u^4<0$,  
and is directed outwards the gain domain: $j>0$ (since $j$ tends to zero as $\eta\to \infty$).  This   means that maxima of $|j|$ are achieved at some  points located inside the gain domain, i.e. $|\eta_*|<\eta_n$. In such points the amplitude of the field is given by 
$
u_*^2=p_iR(\eta_*)/\alpha
$
 (notice that $\eta_*$ itself depends on the gain coefficient).  
 Considering the symmetric  one-hump mode in the case of one gain channel [i.e.   when functions $u(\eta)$ and $j(\eta)$ feature only single maximum],  one has the two maxima of $j_\eta$ at  $\pm \eta_*$, and hence 
 $j_\eta>0$ in the interval $|\eta| <\eta_*$. Therefore the amplitude of the field is bounded by the interval $u_*^2\leq {\cal A}^2\leq p_iR/\alpha$. 

\begin{figure}[h]
   \begin{tabular}{c} 
\includegraphics[width=0.95\columnwidth]{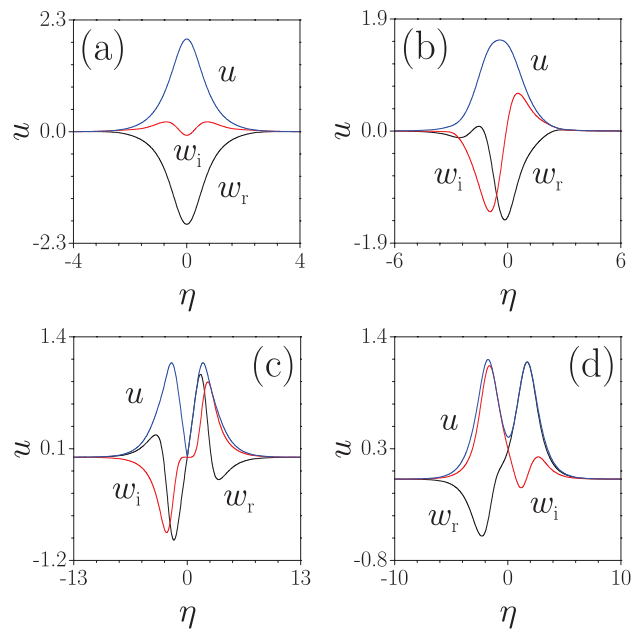}
  \end{tabular}
\caption{(Color online)  Symmetric (a) and asymmetric (b) one-hump solitons at $n=1$, $p_i=3.5$, $\alpha=1.2$, and (c) and (d) respectively profiles of symmetric and asymmetric two-hump solitons at $n=2$, $p_i=2.5$, $\alpha=1.8$. The modes in panels (a) and (b) correspond to circles in Figs.~\ref{fig3}(a) and~\ref{fig3}(b). Hereafter all quantities are plotted in arbitrary dimensionless units. }
\label{fig1}
\end{figure}

The above prediction of symmetric and asymmetric modes 
was confirmed in simulations [Fig.~\ref{fig1}].
We  observed that while the growth of zero background is suppressed at large $\eta$, the light concentrates inside the amplifying channels. Strictly speaking this feature is typical for  symmetric modes. A maximum of an asymmetric mode is shifted from the gain peak  and the width of the mode grows with $p_i$, according to the estimates presented above. This broadening of the soliton   leads to the situation where an appreciable part of the light energy concentrates outside the gain channel for large $p_i$.   

\begin{figure}[h]
  \begin{tabular}{c} 
	\includegraphics[width=0.95\columnwidth]{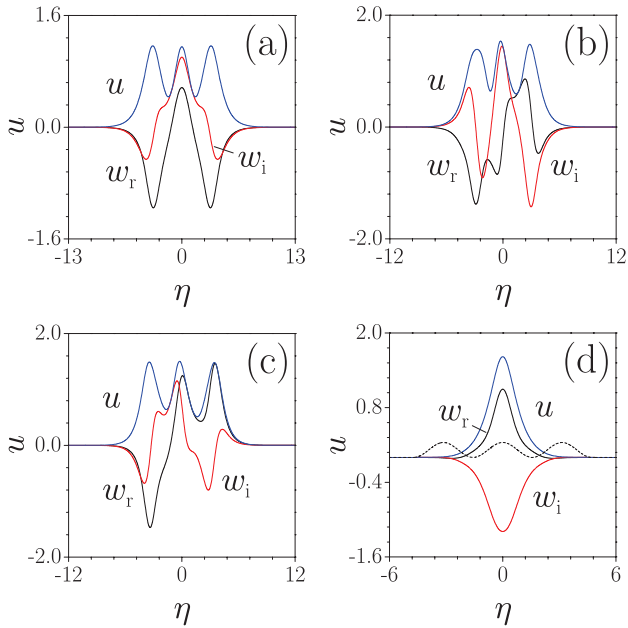}
  \end{tabular}
\caption{(Color online)  Profiles of a symmetric three-hump soliton at  $n=3$, $p_i=2.5$ [it corresponds to the circle in Fig.~\ref{fig3}(c)] (a), and two coexisting asymmetric three-hump solitons at $n=3$, $p_i=4.0$,(b) and (c). (d) One-hump soliton at $n=3$, $p_i=4.0$. The dashed line schematically shows the gain landscape. In all cases  $\alpha=1.8$.}
\label{fig2}
\end{figure}

The observed asymmetry in the field modulus  remains relatively small for all considered $p_i$ (it is most clearly visible in $w_{r,i}$ distributions), but  becomes  more pronounced in systems with larger number of channels [see Fig.~\ref{fig2}].
The number of humps of stable solitons  coincides  with the number of the amplifying channels. In Fig.~\ref{fig2}  we show the   symmetric and two co-existing asymmetric modes for the case of three amplifying channels  (we obtained similar  solitons in landscapes with $n$ up to $20$). 

Asymmetric states in a system where gain landscape is  symmetric and all other parameters are uniform, i.e. the {\em symmetry breaking}, is an 
unexpected result. Indeed, unlike in conservative systems,  the understanding of the phenomenon  cannot be related to the energetic arguments. Our system also does not allow for reduction to a simpler discrete model, as this happens, say, in the case of a double-well potential.
Moreover, in our case the symmetry breaking occurs even for a single gain channel in 
contrast to conservative systems (where at least two potential minima are required).

\begin{figure}[h]
  \begin{tabular}{c} 
	\includegraphics[width=0.9\columnwidth]{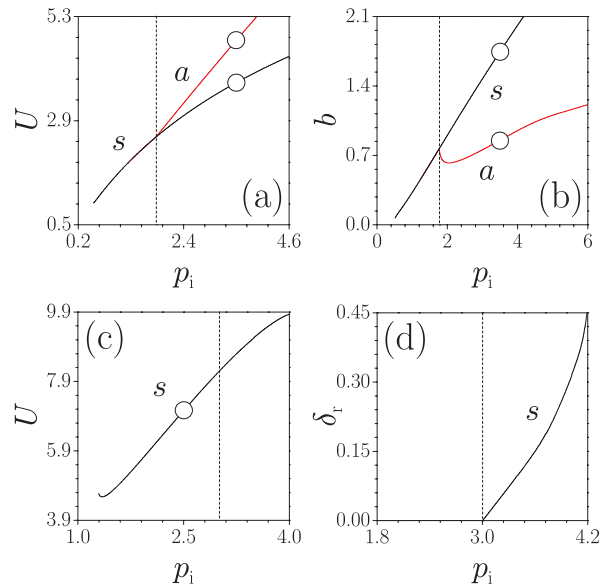}
  \end{tabular}
\caption{(Color online)  Energy flow (a) and propagation constant (b) {\it vs} $p_i$  for symmetric "s" and asymmetric "a"   one-hump solitons at $n=1$,  $\alpha=1.2$. Energy flow (c) and 
  perturbation growth rate (d) {\it vs} $p_i$  for symmetric three-hump solitons at  $n=3$,  $\alpha=1.8$. The  dashed lines 
indicate the borders of stability domains  $p_i=p_i^{cr}$ for symmetric modes.}
\label{fig3}
\end{figure}

We performed numerical study of the whole branches of the solutions and studied their stability [Fig.~\ref{fig3}]. 
In   Fig.~\ref{fig3} (a), (b) for $n=1$ we   observe  two branches of the solutions (notice, the propagation constant $b$ 
is not a free parameter), one of them corresponding to the symmetric solitons, and another one, bifurcating form the symmetric branch at certain value $p_i=p_i^{cr}$, that corresponds to the asymmetric solutions (having smaller amplitudes and larger widths as compared to the symmetric ones). The dependences $U(p_i)$ and $b(p_i)$ for the both branches
well reproduce the estimates presented above.
The linear stability analysis of the modes is performed by plugging in the perturbed  field $\displaystyle{q=\left(w+ve^{i\delta\xi}
 \right)e^{ib\xi}}$ into Eq.~(\ref{NLS}) and performing linearization around $w$.
For odd numbers of amplifying channels,  exactly at the bifurcation point $p_i^{cr}$ the branch of symmetric solutions looses its stability, while the stable asymmetric branch emerges [see the dashed lines in Figs.~\ref{fig3}(a) and (b)]. Since the asymmetric modes appear in pairs (corresponding to the left and right shifts of the  maximum outwards the origin) at the point where the symmetric mode  becomes unstable we deal with the pitchfork bifurcation. 

\begin{figure}[h]
   \begin{tabular}{c} 
\includegraphics[width=0.9\columnwidth]{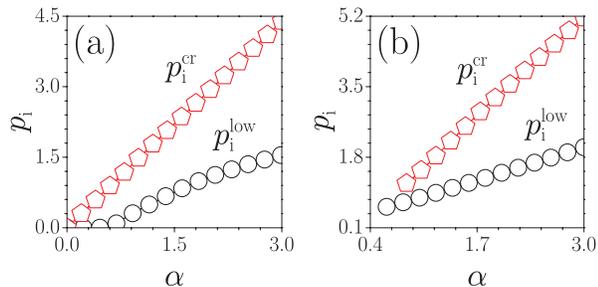}
  \end{tabular}
\caption{(Color online) The domain of existence ($p_i\geq p_i^{low}$)   and stability domain  ($p_i^{cr}\geq p_i\geq p_i^{low}$) for one-hump (a) and three-hump (b) solitons on the plane  ($\alpha,p_i$). For small $\alpha<0.6$ symmetric one-hump solitons can be obtained even when $p_i\to 0$  but for moderate and high nonlinear losses they exist only above certain minimal gain $p_i=p_i^{low}$}
\label{fig4}
\end{figure}

For   small $p_i$    symmetric solitons broaden dramatically and may expand far beyond the region with gain   (there is always a flow of energy outwards amplifying region). Increase of $p_i$   results in growth of the peak amplitude and progressive localization of the soliton inside the amplifying domains.  According to the above estimates   for $n=1$   and  for sufficiently small   $\alpha$  the energy flow and propagation constant of a symmetric soliton  are monotonically increasing functions of  $p_i$ [Figs.~\ref{fig3}(a) and (b)]. Note, that while for  small values of  $\alpha$ the symmetric one-hump solitons can be found even for $p_i\to 0$, for moderate and high alpha values such solitons exist only above certain minimal value of gain coefficient $p_i^{low}$ [see Fig.~\ref{fig4} (a)].

For even $n$ 
the symmetric modes appear  unstable in the whole domain of existence, and the only stable modes are asymmetric ones. In this case the dependencies $U(p_i)$ for symmetric and asymmetric modes do not overlap and no bifurcations occur. Except for stability, 
 other properties of modes supported by even and odd number of amplifying channels are similar. 
Gain 
with multiple amplifying channels also supports solitons with the number of humps smaller than the number of the channels [Fig.~\ref{fig2}(d)].

 In dissipative multi-hump solitons both $w_r$  and $w_i$ change their signs in neighboring channels with gain, while 
 the  
field amplitude $u$ is nonzero even in the regions between the channels.  
This follows from (\ref{rho}). Indeed, let us  assume that at some 
point $\tilde{\eta}$ the field is zero,   i.e. $u(\tilde{\eta})=0$. 
Since
$u(\eta)$ is nonnegative, in the vicinity of $\tilde{\eta}$ we have:   $u(\eta)={\cal O}\left((\eta-\tilde{\eta})^2\right)$ and $j_\eta={\cal O}\left((\eta-\tilde{\eta})^4\right)$. Expanding $u(\eta)$ and $j(\eta)$ in the Taylor series in the vicinity of $\tilde{\eta}$ we find subsequently that all the expansion coefficients are zero, what means that if $u$ becomes zero at some point,
$u(\eta)\equiv 0$ and $j(\eta)\equiv 0$. 

   The critical gain, at which the bifurcation 
  occurs, increases almost linearly with 
  $\alpha$, so that the domain of stability of symmetric solitons $p_i^{cr}\geq p_i\geq p_i^{low}$   expands with $\alpha$ [Fig.~\ref{fig4}(a)]. In the case of symmetric multi-hump solitons the energy flow grows with  $p_i$ monotonically, except for the narrow region close to the threshold  $p_i= p_i^{low}$ below which no multi-hump solitons can be found [Fig.~\ref{fig3}(c)]. Multi-hump solitons are stable in the region adjacent to $p_i^{low}$, but  
  increase of $p_i$ results in their destabilization 
  at $p_i=p_i^{cr}$ [dashed line in Fig.~\ref{fig3}(c)]. A typical dependence of perturbation growth rate $\delta_{r}$ on gain parameter for $n=3$ is shown in Fig.~\ref{fig3}(d). The stability domain of symmetric three-hump soliton expands almost linearly with increase of $\alpha$ [Fig.~\ref{fig4}(b)]. 
 
Destabilization of symmetric multi-hump states is accompanied by the appearance of several stable branches of asymmetric multi-hump solitons. Thus asymmetric modes depicted in Figs. ~\ref{fig2}(b) and~\ref{fig2}(c) that are both stable and corresponding symmetric unstable three-hump mode (not shown) coexist for the same values of $p_i,\alpha$.  

With increase of the number of the gain channels  the picture becomes even richer. When the number of channels is odd, the number of asymmetric modes that can be  stable all together  for fixed $p_i$ and $\alpha$ values increases.  This feature indicates on the presence of several stable attractors (multistability) in multichannel 
landscapes.
The critical value of the gain coefficient (i.e. the bifurcation point)  $p_{i}^{cr}$ also grows with $n$, reaching however certain saturation value. In particular, for $\alpha=1.5$ this  value is about $2.59$ and  it is reached already at $n=7$.

Summarizing, we  reported   the symmetry breaking of dissipative soliton supported by a single or multiple amplifying channels embedded in the cubic medium with nonlinear losses, which occurs through the loss of the stability of the symmetric family at a point of the  pitchfork bifurcation.  Since inside the stability domains 
solitons are attractors with sufficiently large basin they can be excited with a variety of regular or noisy input patterns. 
For Gaussian inputs and single amplifying channel stationary  solitons may form already after propagation over 20-30 diffraction lengths. 
In the case of multiple gain channels the bifurcation leads to appearance of 
several stable asymetric modes.  While the bifurcation type  is  the same as one leading to appearance of the self-trapped states in a conservative double well potential~\cite{Panos}, here we deal with pure dissipative phenomenon and the system does not possesses any characteristic scale related to its conservative part. 


\end{document}